# High-performance Decoder for Convolutional Code with Deep Neural Network

Jiang Xiaobo, Zhang Fang and Zeng Zhen

School of Electroinc and Information Engineering

South China University of Technology

Email: jiangxb@scut.edu.cn, xbjiang@gmail.com

*Abstract:* The use of deep neural network for decoding error control code will encounter two problems, namely, the high-precision requirements of the error control code and the complexity of the neural network due to the long code. In this paper, a deep neural network decoder is proposed to solve the decoding problem of long code by using the nature of convolutional code window decoding. A deep neural network decoder is utilized as a weak classifier, and an integrated decoder is proposed to improve the decoding performance greatly. The Viterbi decoder is improved by approximately 2 db at a bit error rate of $10^{-5}$. Both decoder methods proposed in this paper can be decoded in parallel and are suitable for high-bit-rate applications. This study reveals that the accuracy of neural networks can reach $10^{-8}$ or more.

## I. INTRODUCTION

Deep learning has attained remarkable progress in recent years and has been widely applied to computer vision [1], natural language processing [2], autonomous driving [3], and other fields, with impressive results. Although communication is still developing at a high speed [4][5][6], deep learning technology has been adopted and it is expected to achieve new breakthroughs. Data are the fuel for deep learning. The error control code can provide a large amount of data beneficial for adopting a deep learning method. However, the error control code contains an error rate of $10^{-5}$ or less, which considerably exceeds the accuracy of the current deep learning application. The average application only requires 90% accuracy. Thus, the application of deep learning method to the field of error control code not only introduces new decoding methods for error control codes but also further investigates deep learning capability and provides a method for evaluating the capabilities of deep neural networks.

Neural networks have long been applied to error control codes [7][8][9][10]. The deep learning method has achieved progress in the field of error control codes[11][12][13]. Karami et al. used the perceptual neural network to decode the LDPC code [14]. Their approach can solve the long LDPC code, but its performance is much more serious than that of the BP algorithm. Tobias[15] utilized deep neural networks to solve short polarization codes. The performance of these networks is close to that of the MAP algorithm. Eliya et al. used deep neural networks to solve the BCH code [16]. They initially proposed an information transfer method to solve the BCH code and then improved the method through a deep learning approach. However, the performance of proposed algorithm and original BCH were no compared. Deep neural networks are also applied to linear block codes] [17][18][19].The current report and our own research indicate that deep learning is outstanding in short codes. The excellent classification capability of deep learning is reflected in the decoding performance and can approach or exceed the traditional decoding method by directly adopting channel output. However, the complexity of constructing neural networks through this method increases sharply with the increase in code length, thereby affecting its feasibility. Long code generally considers the code structure to construct a neural network, and a considerable gap still exists between performance and traditional methods. Existing neural network error control codes have not been reported in convolutional codes.

In this paper, a decoding method of deep neural network for convolutional codes is proposed. This method utilizes the characteristics of convolutional code window decoding to solve the problem of long code decoding and achieve the performance of Viterbi decoding. Furthermore, a decoding method of integrated neural network for convolutional codes is proposed along with the integration method, which greatly improves the performance of neural network decoding. The decoding method is improved by approximately 2.5 db compared with Viterbi decoding at a bit error rate of $10^{-5}$.

This paper is arranged as follows. The second part introduces the system design and data sets used in the research. The third part provides the structure and test method of neural network decoding. The fourth part discusses the integrated decoding method. The fifth part presents the simulation results.

## II. SYSTEM

### A. System Framework

Figure 1 shows the system framework of this study. At the transmitting end, the information bit X of length l is encoded by a convolutional code to become a codeword of code length N. In this study, a convolutional code (2, 1, 2) is used, and the generator polynomial is provided by Equations (2-1) and (2-2). The coded codeword is digitally modulated with BPSK. The modulated symbol enters the receiving end after adding the white noise channel. The receiving end receives the noisy information and demodulates to obtain the maximum likelihood information LLR. Then, this information enters the neural network decoder. The decoding result is obtained by neural network decoding and integrated decoding.

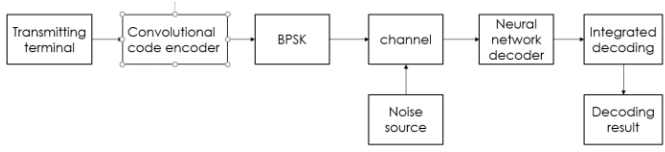

Fig. 1 System frame

$$g1 = 1 + x + x^2, \qquad (1)$$
$$g2 = 1 + x^2. \qquad (2)$$

Two neural network decoders are proposed in this study. This first one directly uses a deep neural network decoder, and the other utilizes the first decoder as a sub-decoder and applies voting to form an integrated decoder to obtain the final result.

### B. Data set

This study adopts a multi-class hypothesis to construct a deep neural network decoder. Figure 1 shows the system used in building a data set. The training data window size was set to 2 w, and the received LLR information codeword was divided into segments. The window moves 2 bits each time (equivalent to 1 bit of information bit change), and the corresponding w bit information is used as the label. The entire data set was combined as a training set.

In this study, L = 10000, and the first 8 bits can be set to 00000000 to facilitate the subsequent shift decoding, such that the noisy codeword is constructed as the w*L*w data set. The training set is disrupted as an experiment in this study. A huge sequence of product code information is obtained by adding 1 db to 8 db different noise volumes. Similarly, L = 10000, and the validation and test sets are constructed. When the training set is not the same, disturbing the data set is not needed.

The original information bit is unsuitable as a label because the multi-classification method is adopted. Thus, the original information bit must be converted into a one-hot code. The input data of the neural network are assumed to be 8 bits per line, that is, 4 information bits, and 16 bits are converted into a one-hot code form. For example, the information sequence is "0000," and the unique heat code is represented as "1000000000000000."

### III. NEURAL NETWORK DECODER

#### A. Neural network structure

A fully connected neural network is used in this study (Figure 2). Table 2 shows the network layer parameters. The constructed neural network structure is the code length of the input convolutional code. In this experiment, the corresponding 16 bits are included, and the hidden n-layer is included in the middle. In this study, n is 1–3. The information bit corresponding to the output layer has a tag of 256 bits, the activation function of the middle layer utilizes the ReLU and dropout layers, and the loss function uses cross entropy.

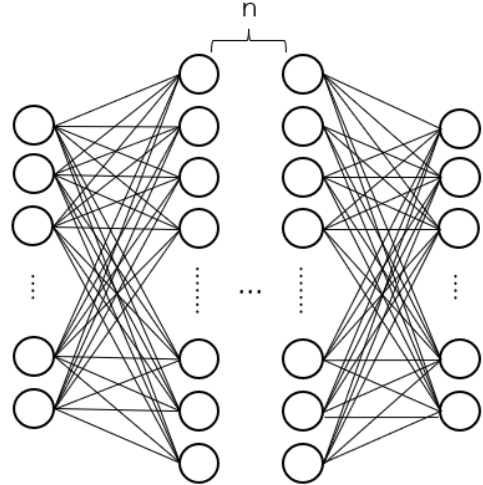

Fig. 2 Neural network structure

| Hidden layer n | Network parameters | Activation Function |
|---|---|---|
| 1 | 512 × 128 | ReLU |
| 2 | 512 × 256  256 × 128 | ReLU |
| 3 | 512 × 256  256 × 128  128 × 128 | ReLU |

Table 2 Neural network layer's parameters

#### B. Parallelism

The neural network decoder proposed in this study adopts a sliding window mode that can realize parallel decoders, thereby improving throughput.

The original information bit length in establishing the input data sample of the neural network decoder for the (2, 1, 2) convolutional code encoding sequence is 1, the encoded noise-added codeword length is 2 l, and the window length is w. The noisy codeword is cut into 2 l/w parts in accordance with the window length w, thereby forming a matrix similar to 2 l/w×w (as the Fig3 shows).

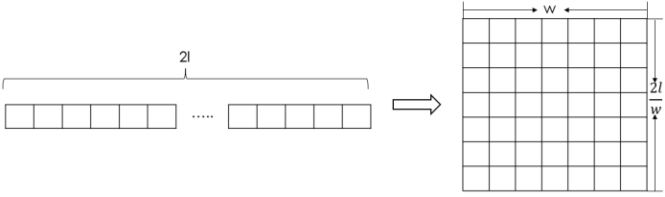

Fig. 3 Construction of data set in direct decoding

The processed data set is input into the neural network, such that the output of the neural network is equivalent to decoding each line of encoded information, and the neural network result is the decoding result of each line of encoded information, thereby achieving parallel operation.

*C. Training and results*

The classification error is minimized in the training. Moreover the batch gradient descent method is used in training to reduce the classification error. Feedforward calculation and backward propagation are utilized to update the weight to train the deep neural network, such that the model possesses strong multi-classification capability. The training and test sets for neural network training are constructed, as described in Section 2.2. The number of training iterations using average stochastic gradient descent algorithm is 10,000, the initial learning rate is 0.01, and the batch size of the training set is set to 128. The number of hidden layers of the neural network can be 1 to 3. The activation function selects the ReLU function.

$$\text{ReLU}(x) = \begin{cases} x \text{ if } x > 0 \\ 0 \text{ if } x \leq 0 \end{cases}.$$

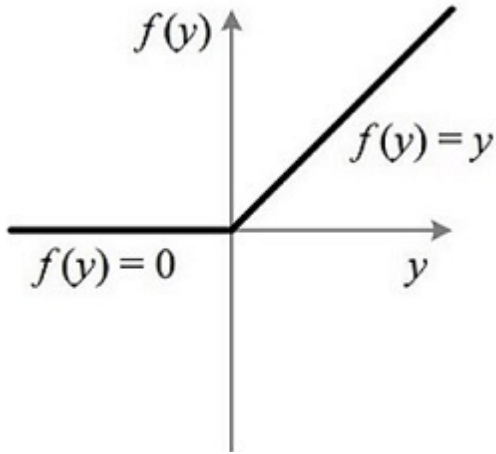

Fig. 4 ReLU function

Figure 5 shows a gap between the performances of direct neural network decoding and Viterbi decoding. From the depth of one-layer full connection to three-layer full connection, the performance of two-layer fully connected neural network decoding is slightly better. However, its performance is still less than veterbi decoder. This phenomenon is expected because the information extraction characteristics are not enough, and a simple fully connected neural network acts as a weak classifier. To improve the performance further, the idea of integration can be adopted.

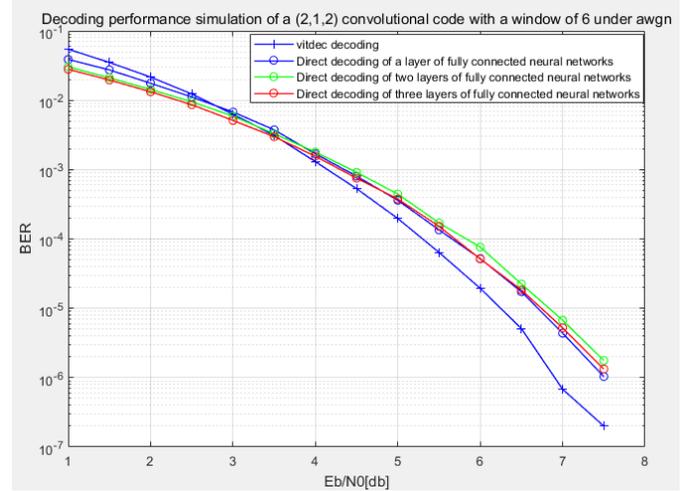

Fig. 5 Performance chart of direct neural network decoding

## IV. INTEGRATED DECODING

A neural network decoder is simply used, and its performance does not exceeds the Viterbi decoder. To exceed the Viterbi decoder performance, simply increasing the network depth or window width is unrealistic. Increasing the network depth may increase network complexity and cause over-fitting. Furthermore, the window width is subject to convolutional codes. Related parameters have restrictions. If the idea of integration is adopted, then different weak classifiers are combined to form a strong classifier, which can improve the decoding performance.

Bagging in the integrated algorithm of machine learning is a method for improving the accuracy of learning algorithm classification. The basic idea is to develop a weak learning algorithm and a training set. As the accuracy of a single weak learning algorithm is not high, the learning algorithm is used to obtain the result samples multiple times for voting, and the prediction accuracy of the learning result is improved.

In this study, we use bagging to input the neural network after constructing the data set. The decoding result of the single coding field is divided by the output result of the neural network, and the final decoded result is selected by voting on the decoding result.

*A. Data Processing*

Multiple decoding results corresponding to a single coded bit should be obtained using the bagging concept to integrate decoding. In this paper, the codeword length is 2l, the data construction window is w, and each mobile step is 2 bits (one bit information bit), as shown in FIG. 6. Then the same bit information will appear in w/2 data, as shown in Figure 7.

Thus we get w/2 decoding results for each single encoded information bit.

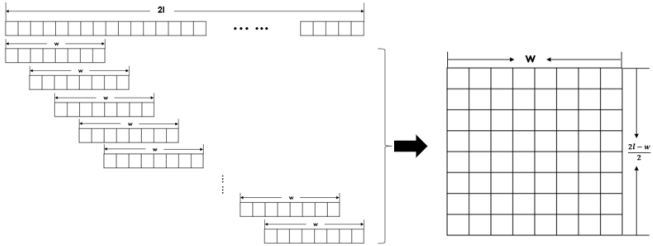

Fig. 6 Process diagram for data set construction

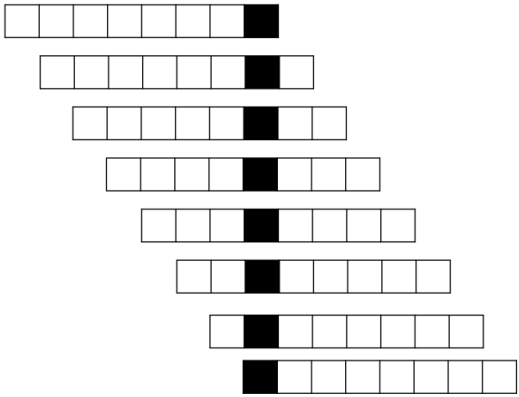

Fig. 7 Sample chart of the data set

Figure 7 shows eight decoding results after decoding each encoded noise information by using the neural network decoder. Eight weak classifiers exist for the same bit information. The decoding performance can be improved by voting on these eight decoding results.

The decoded data set can be obtained by inputting the processed data set into the neural network decoder. The decoding result is voted in accordance with the flowchart shown in Figure 6. Finally, our decoding result is obtained. In Figure 7, the same color decoding block represents the result of decoding the same code field. After this process, w decoding result samples are obtained. When processing the data set sample results, the input neural network output result is similar to the one-hot code form, which is converted into a decimal and then binary result corresponding to the original information sequence.

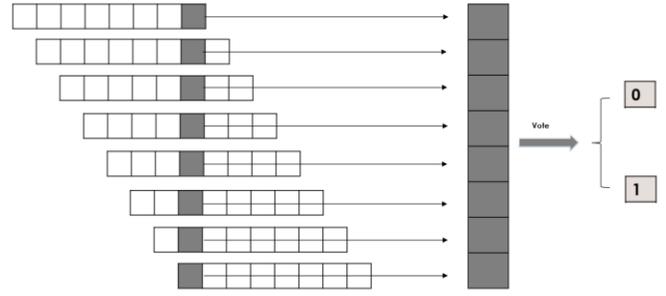

Fig. 8 Voting flow chart in the neural network decoding sample

### B. Decoding voting process

First, the trained deep neural network is decoded. The obtained noisy soft information is input into the neural network in accordance with the input layer size of the deep neural network, and the obtained neural network is output. Second, the integrated method is used, because the first eight bits of the information bit sequence in the previously constructed data set are 0. Then, from the first line of the decoded sample, the decoded output of the first code segment sample through the deep neural network takes the last one. Bit is equivalent to decoding the first information bit once. At this time, the next code segment sample is decoded, because it is a noisy codeword and then sliding n0 bit, which is equivalent to the information being moved. If the information bit sliding distance is 1 bit, then the last bit of the first code segment sample output corresponds to the second last bit decoding result of the second code segment sample and to the same information bit on the convolutional code. When the number of slides reaches the output layer size n of the deep neural network, we obtain n weak classifiers, including the information bits. Moreover, the n weak classifiers categorize the information bits in different dimensions by using integrated method votes for this information bit, which can result in a strong classifier that yields improved decoding performance.

Figure 7 shows the entire process and location of the decoded bits in different code segment samples. This step is repeated for subsequent codewords, thereby fully decoding the convolutional code. For the beginning of the convolutional code sequence, the number of (N-n0) zero bits can be added before the codeword at the time of decoding, such that the decoding is guaranteed to be decoded from the beginning of the original convolutional code. The last codeword of the code sequence can be added with (N-n0) number of zero information bits at the end of the information bit during decoding. Thus, the input requirements of the deep neural network are met, and the calculation process of deep neural network decoding is continued.

### C. Parallel decoding method

Parallel decoding can greatly reduce the decoding delay because the code length of the convolutional code is often long in actual use. Each parallel process is indistinguishable from the original. The original serial head and tail need to join a part of 0, and the starting position of each parallel process needs to be connected to the tail of a parallel process, except for the start and end. The tail of each parallel process needs to link the head of the next parallel process, which is not involved in decoding.

Figure 9 shows that each process only translates the codewords associated with itself. Using a codeword with a length of 14 is only necessary before the codeword at the entrance of each parallel decoding, and the last two is the codeword at the decoding entry, thereby forming a length of 16 codewords as the input of the deep neural network for translation. The code outputs the information bits of the corresponding codeword at the decoding entry. Similarly, the shift begins, and the next codeword is decoded. The decoding process is the same as before. The previous 14 codewords belong to the decoding content of another parallel decoder and do not participate in the decoding of this parallel process. The parallel decoding method does not degrade the decoding performance, the position of the parallel decoding entry on the codeword has no requirement, and multiple parallel processes can be implemented.

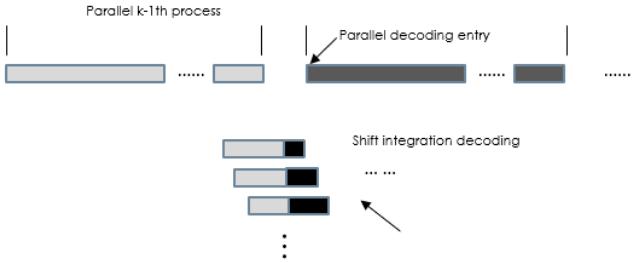

Fig. 9 Parallel decoding process

## V. SIMULATION RESULTS

In this experiment, Python and MATLAB were used to verify the experimental simulation. Deep neural network integrated decoding was compared with Viterbi decoding. Figure 10 shows the simulation results.

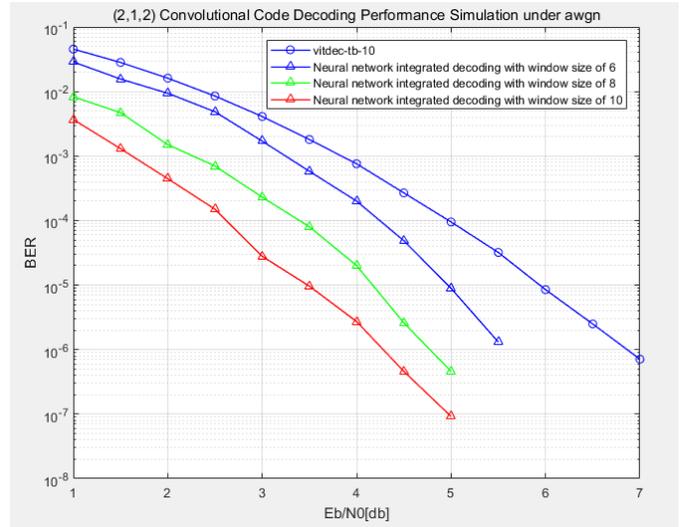

Fig. 10 Integrated shift decoding performance map

Simulation results show that the performance of integrated decoding is greatly improved compared with that of Viterbi decoding. Under the $10^{-5}$ condition, the integrated decoding performance at the window of 10 is approximately 2.5 db higher than that of Viterbi decoding. The integrated decoding performance increases as the window increases.

## VI. 6 CONCLUSION

This study proposes a neural network decoding method for convolutional codes. One is a fully connected deep neural network decoding method, and its decoding performance is comparable with that of the Viterbi decoding method. An integrated shift decoding method is proposed by using the deep neural network decoder as a weak classifier, thereby greatly improving the performance. The performance of integrated shift decoding is improved by 2.5 db compared with Viterbi decoding. The wider the window of the weak classifier, the better the performance. Both decoders proposed in this paper can use parallel decoding, which is suitable for the throughput application requirements.

In this paper, a neural network for solving the problem of error control code length is proposed. Through our research, we can find that the accuracy of deep neural network can reach more than $10^{-8}$. This study also introduces a test method for the classification capability of deep neural networks. The next step is to investigate the hardware implementation of the neural network decoder and research on how to simplify the neural network and reduce the complexity of hardware implementation while maintaining high performance. The principle of neural network error correction control code is also worth studying. Neural network decoding involves classification, and noise reduction or the derivation of decoding formulas is an interesting topic.

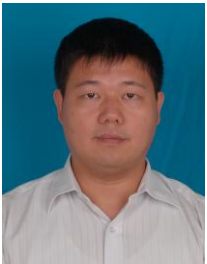
**X.B. Jiang**, received the bachelor's and master's degree from Zhejiang University in 1994 and 1997, and the PhD degree from institute of microelectronics, the Chinese academy of sciences in 2004. After graduation, he joined the School of information and engineering, South China University of Technology. His research interests include flexible electronics,low-power IC design and communication baseband chip design. Email: jiangxb @scut.edu.cn.